\documentclass[12pt]{article}
\usepackage{graphicx}
\usepackage{cite}

\begin{document}

\title{\bf Analysis of the Accuracy of Prediction\\ of the Celestial Pole Motion}
\author{Zinovy Malkin, Pulkovo Observatory, malkin@gao.spb.ru\\[1em]
\small Received April 14, 2010; in final form, July 08, 2010}
\date{~}
\maketitle

\begin{abstract}
VLBI observations carried out by global networks provide the most accurate values of the
precession-nutation angles determining the position of the celestial pole; as a rule, these results become
available two to four weeks after the observations. Therefore, numerous applications, such as satellite
navigation systems, operational determination of Universal Time, and space navigation, use predictions of
the coordinates of the celestial pole. In connection with this, the accuracy of predictions of the precession-
nutation angles based on observational data obtained over the last three years is analyzed for the first time,
using three empiric nutation models---namely, those developed at the US Naval Observatory, the Paris
Observatory, and the Pulkovo Observatory. This analysis shows that the last model has the best of accuracy
in predicting the coordinates of the celestial pole. The rms error for a one-month prediction proposed by this
model is below 100 microarcsecond.
\end{abstract}

\section{Introduction}

A prediction of the Earth orientation parameters (EOP)---the coordinates of the Earth's pole, Universal
 Time, and the coordinates of the celestial pole--- is an extrapolation of EOP observations over a given
time interval (the prediction length) following the last date of observations. Such predictions are necessary
for numerous scientific and practical applications, and numerous studies (see, for example, [1-6] and references therein)
have been devoted to improvements in methods and analyses of accuracies of such predictions.
Almost all studies in this field are concerned with predicting the coordinates of the Earth's pole and Universal Time.

The reason for this is that, in contrast to these, precession-nutation models can be constructed for
the motion of the celestial pole (CP), with the accuracies of these models satisfying those required for
numerous practical applications. In this situation, improvements in the accuracy of the precession-nutation theory simultaneously ensures improvements
in the predicting accuracy. The last theory adopted by the International Astronomical Union (IAU) as the international standard, which incorporates
the nutation theory IAU2000A (the IAU resolution B1.6 2000) and the precession theory P03 (the IAU resolution B1 2006), has an accuracy
of 100 microarcsecond ($\mu$as). However, this accuracy does not reflect the real error in the CP position calculated
 by the IAU theory, since the latter describes only part of the total motion of the Earth's rotational axis.
In addition to the terms described by the theory, this motion also includes a number of terms which cannot
be modeled at present, since they involve certain unpredictable effects, with the total contribution of
these reaching 400~$\mu$as. The most important of these effects are the free core nutation (FCN) and the trend.

Thus, applications, which require only about 500~$\mu$as accuracy in the precession-nutation modeling
 do not require any special predictions. However, applications with higher requirements for modeling
of the real-time CP motion, and all the moreso the CP motion prediction in the future, require modeling
and predicting of the difference between the real CP motion and the IAU theory, called the celestial
pole offset (CPO). The CPO values $dX$ and $dY$ are determined by Very Long Baseline Interferometry
(VLBI) observations and the EOP services. There are several empiric models in use, which are continuously
supported and can be used in practice for these purposes [7].

For each model, a CPO series is regularly published for each given time together with a prediction.
However, only a few studies have considered the errors of the CPO predictions [8-10], and these studies
have used historic data on the CP motion, which can distort the prediction accuracies, especially for short-
term predictions [11]. In addition, only the study of Lambert [10] considers a model that is available for
practical applications. Thus, users of predictions of the CP motion have no opportunity to obtain statistically based estimates of the associated errors.
The only known attempt to estimate the real error of a CPO prediction was undertaken by the International Earth Rotation and Reference Frames Service
(IERS) [12]. Unfortunately, this estimate analyzes only two predictions using unaccessible data and an unknown model. The graphical data presented in [12]
indicate a CPO prediction accuracy of about 150~$\mu$as for prediction lengths of less than one year.

Our present study analyzes real CPO predictions made by various EOP services in 2007-2009, using
models saved in our database. These predictions are compared with the final CPO values determined by
the International VLBI Service for Geodesy and Astrometry (IVS) using VLBI observations carried out
by global networks [13, 14]. The statistical processing of these data provides reliable estimates for the quality
of the CP prediction methods presently used. These estimates are calculated for the rms error, maximum
error, and prediction efficiency. The first two quantities characterize the absolute error of the prediction,
while the last estimates the relative error.

\section{Celestial pole offset determined from VLBI observations}

It is well known that VLBI carried out on global networks provides the most accurate data on the
CP motion. These IVS-coordinated observations require  the participation of many observatories located
in various countries [13]. In general, high-accuracy EOP data can be obtained from any set of VLBI
stations with the necessary geometric parameters, most importantly, their size [15]. Since the data processing
 is very labor-intensive (we will analyze this below), VLBI observations become available with a
delay reaching several months, which is insufficient for practical EOP applications, although the VLBI
method is fundamental for determining the CP position  and Universal Time. To improve the immediacy
of the EOP monitoring, IVS observations are incorporated  in two special programs called R1 (as a rule,
carried out from Monday 17$^h$ UTC to Tuesday 17$^h$ UTC) and R4 (as a rule, carried out from Thursday
18$^h$30$^m$ UTC to Friday 18$^h$30$^m$ UTC). The observations incorporated in these programs have priority
at all stages of the processing, reducing the delay in obtaining the EOP as much as possible. In addition,
these observations are optimally scheduled to reach a higher EOP accuracy.

The main principle of VLBI is that each station (radio telescope) of the VLBI network detects the
radio-source signal independently. The resulting data  are then sent to a correlator for processing. The
correlation results include the interference delays and frequencies recorded in the form of IVS Mark-3 DBH
(database v.~1 in the international terminology) standard  files for the S and X bands. Further is the preliminary
 processing of the data, which takes into account the influence of the ionosphere, as well as data on
the cable delay and meteorological and some other parameters both immediately detected at the stations
and calculated theoretically. This results in final observational  files (database v.~3 or 4 in the international
terminology), which are used by the IVS centers for their calculations of the Earth rotation parameters.
After processing the observations, the individual centers  calculate IVS combined EOP series. All this data
processing contributes to the time delay before the CPO data are generally available. Here, the time delay
is the difference between the date when the EOP data become available and the epoch when the data are taken.

Let us consider the approximately distribution of this time delay over the various processing steps for
the EOP data determined in the IVS R1 and R4 programs.

\begin{enumerate}
\item Data delivery to the correlator. Most VLBI observations carried out on global networks are
recorded on magnetic disks and delivered to the correlator by express mail, which usually takes from
three to seven days. This usually delays the start of the correlation processing by six to eight days.
\item Data correlation. The correlation usually takes from two to five days. This also includes the preliminary
analysis and data calibration, as well as the formation of the final observational file and its installation
on the IVS servers for subsequent use at all processing centers.
\item EOP calculations at the IVS processing centers. As a rule, first results become available at the
centers forming the final observational files almost without any delay. At the other data-processing centers,
the EOP data usually become available in one to two days\footnote{http://vlbi.geod.uni-bonn.de/IVS-AC/data/timeliness\_2.html}.
\item Calculation of the IVS combined EOP series. This usually takes from two to ten days.
\end{enumerate}

Our monitoring shows that the minimum delay for the IVS EOP series in 2009 is 12 days and the maximum
 37 days, with the median delay being 19 days. Thus, the CPO data of individual processing centers
usually become openly available 1.5-2 weeks after the observations, while the IVS EOP series become
available 2-4 weeks after the observations.

The time delays for the observational data obtained in other IVS programs range from several weeks to
several months, although these are almost insignificant for the final CPO data. Consequently, a CPO
prediction for five weeks satisfies most practical applications. However, some other practical and research
applications require accurate CPO predictions over longer time intervals.

\section{CPO models used}

To analyze the accuracy of the CP motion predictions, we used models freely accessible for research
and practical applications. There are two types of such models, namely, the FCN models and the CPO
models, which are related as follows:
\begin{equation}
CPO = FCN + \sum T_i + \sum P_j \,,
\label{eq:cpo-fcn}
\end{equation}
where $\sum T_i$ is the sum of trends, for example, resulting  from errors in the precession model, and $\sum P_j$
is the sum of (quasi)harmonic terms, for example,  resulting from errors in the nutation model or
geophysical processes. The FCN models are quasi-harmonic with variable amplitudes and phases [7],
and their maximum amplitude reaches about 350~$\mu$as. At present, the contribution of the second term
reaches 200~$\mu$as and is continuously increasing; the contribution of the third term is probably below
100~$\mu$as. We have already noted that the sum of the CP coordinates determined by both the IAU
precession-nutation theory and the celestial pole offset provides the most accurate data on the CP
motion. Using the FCN models instead of the CPO models results in appreciable errors in the CP position,
with these errors increasing with time.

At present, a search for the most accurate CP coordinates and their predictions finds the following three CPO options.
\begin{description}
\item{\it The NEOS CPO series.} This series is calculated at the US Naval Observatory in the framework of the
national US EOP service, and is an official combined online IERS EOP solution that includes, among the
other data, the CPO series with a 90-day prediction. The algorithm used calculates the summary series of
the CP coordinates using several selected VLBI EOP series and compares the results with the precession-
nutation model [16, p. 73]. Formally, the CPO series is updated everyday; in fact, the new data become
available when each new VLBI session is processed, i.e., two to three times per week, on average.
\item{\it The SL (S. Lambert) FCN model.}  This model is calculated at the Paris Observatory and is recommended
by the IERS Conventions (2003) which collect the most up-to-date astronomical and geophysical
models used for the reduction of astrometric observations [17, Ch. 5]. The combined IERS EOP
series calculated at the Paris Observatory and the analysis of its deviation from the precession-nutation
theory is the basis for the FCN calculations. The FCN parameters are calculated via least-squares fits in
sliding two-year intervals, with the mean CPO values being removed within each interval [10]. In contrast to
other series, this series contains no trends. The FCN parameters obtained within the last interval are used
for a one-year prediction. Since 2009 the SL series is updated once per year---on July 1 (Lambert, private
communication).
\item{\it The ZM2 CPO series proposed by the author.} This CPO series is calculated at the Pulkovo Observatory
using the IVS CPO series [7], and includes a two-year prediction. The construction of this series is
described below. Starting from 2008, this series has been formally updated every day (it was updated about
once per month before 2008), however, in practice, the new data become available only after the IVS series
has been updated.
\end{description}

The Table~\ref{tab:models} presents the main characteristics of the given models.

\begin{table*}
\centering
\small
\caption{Main CPO model characteristics.}
\label{tab:models}
\begin{tabular}{lcccl}
\hline
Model  & Type& Forecast & Smoothing & \multicolumn{1}{c}{URL} \\
\hline
USNO   & CPO & 3 months & weak      & http://www.usno.navy.mil/USNO/earth-orientation \\
SL     & FCN & 1 year   & strong    & http://syrte.obspm.fr/$\tilde{\phantom{a}}$lambert/fcn \\
ZM2    & CPO & 2 years  & moderate  & http://www.gao.spb.ru/english/as/persac  \\
\hline
\end{tabular}
\end{table*}

The ZM2 model is very simple and is calculated using Gaussian smoothing (high-frequency Gaussian
filtering) of the IVS series individually for $dX$ and $dY$:
\begin{equation}
x^*(t) = \frac{\sum_1^n p_i q_i x_i}{\sum_1^n p_i q_i}\,, \quad q_i = \exp[-(t-t_i)^2/2a^2]\,,
\label{eq:gauss}
\end{equation}
where $x_i$ are the original data containing $n$ points detected with weights $p_i$ at the time $t_i$, $x^*(t)$ are the
smoothed $x$ values for the time $t$, and $a$ is a smoothing parameter. When $a$ increases, the smoothing also increases,
i.e. the final series becomes smoother. Since the time $t$ is arbitrary, this method can also be used for
interpolation; see below. This feature of the Gaussian smoothing, as well as its insensitivity to coincident
or very close ti, make this method preferable over the Whittaker and Vondra\'k methods frequently used
in astrometry [18]. Autoregression fitting is used to provide the prediction for the ZM2 series.

Figure 1 compares the three models with the IVS CPO series, and Fig.~2 presents the differences
between these. The NEOS and ZM2 models are systematically close to each other, since these are
constructed similarly, although the ZM2 data are smoother, as seems to be more natural. We suggest
that the NEOS model is distorted by observational noise. Some systematic differences between these
two models can probably be explained by differences in the VLBI data used, which have been mentioned
above. We note also that the predictions obtained by these two models are very different (see the data
shown in Fig.~1 after the last IVS point).

\begin{figure}[ht]
\centering
\includegraphics[width=0.6\columnwidth,clip]{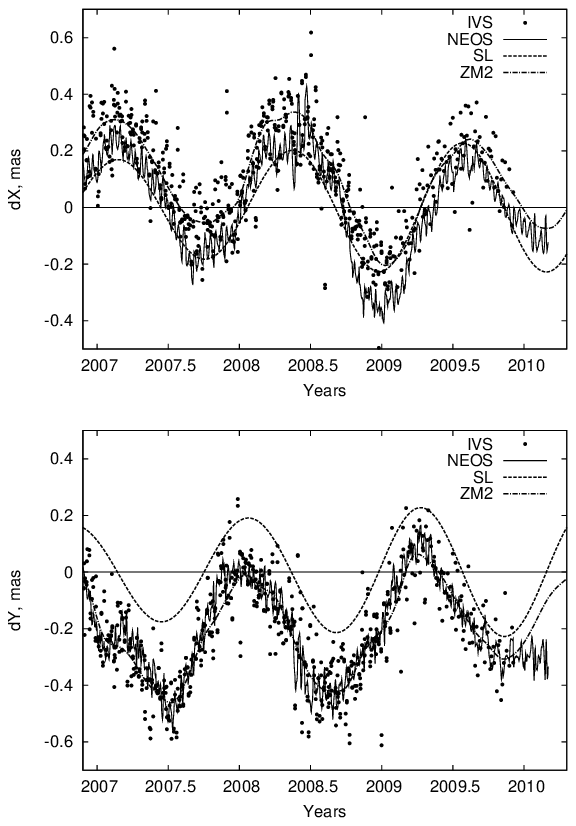}
\caption{The NEOS, SL, and ZM2 CPO models compared with the IVS data.}
\label{fig:cpo_ivs}
\end{figure}

\begin{figure*}[ht]
\centering
\includegraphics[width=\textwidth,clip]{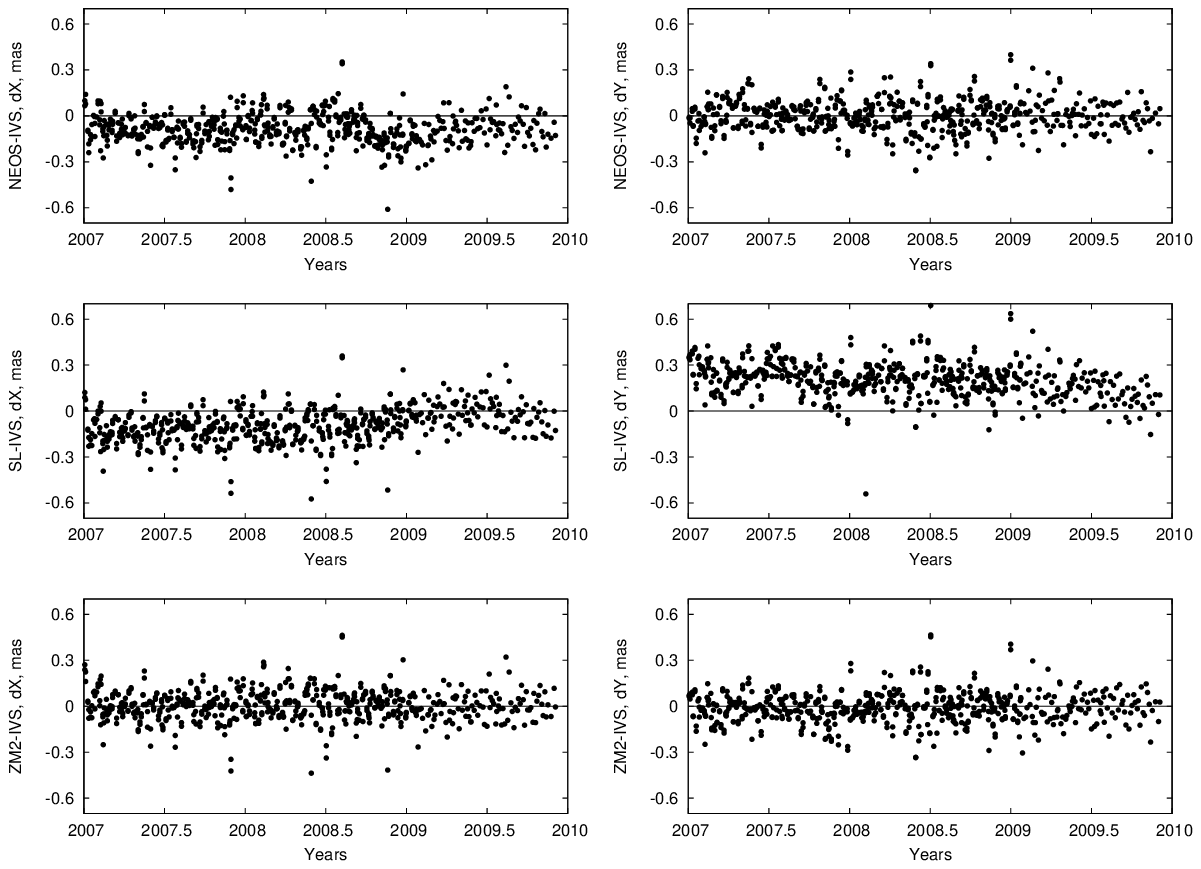}
\caption{Difference between the CPO models and the IVS data ($dX$ is at the left, $dY$ at the right; the NEOS, SL, and
ZM2 models are shown in descending order).}
\label{fig:cpo-ivs}
\end{figure*}

In contrast to CPO models, the FCN SL model includes only the period of 430.2 days [10]. Therefore,
in Fig.~1, the SL model is centered relative to the vertical axis, and shows a substantial difference
between the model and observations (the IVS data). In addition, this model seems to be too smooth, and
does not describe the Earth's rotation in detail.

\section{CPO prediction errors}

To estimate the accuracy of the CPO predictions, we used the predictions made at the US Naval Observatory
(the NEOS model), Paris Observatory (the SL model), and Pulkovo Observatory (the ZM2 model)
in 2007-2009, i.e., over the last three years. For the NEOS and ZM2 models updated daily, we used a
one-week interval for sampling the data. Thus, we used 153 NEOS, 121 ZM2, and 18 SL predictions.

Every study of prediction errors includes an a posteriori analysis of the differences between the
predictions and the final (observational) data. In our case, it is not entirely quite clear what data should be
considered as final; both the CPO series obtained at the individual centers and the combined IVS or IERS
EOP series can be used for this purpose. We decided to take the combined IVS EOP series as the most
appropriate final data, since this series uses all the available VLBI data and avoids possible errors that
arise in the construction of the combined IERS series.

The large amount of noise in the observational data is the other difficulty in the analysis of prediction
errors. This is true both for the individual and the combined CPO series, as we can see in Fig. 1. For
the IVS series, the noise component of the CPO amplitude $\sqrt{dX^2+dY^2}$ estimated as the weighted
Allan variation [19] reaches 118~$\mu$as for the total series obtained over 1984.0-2009.9 and 91~$\mu$as for
the series over the last three years, i.e., for the data used here.

Since the IVS data are given for the mean time of the daily observations, whereas all predictions are
given for the beginning of the Greenwich day, we must reduce the IVS series to a standard smooth
equidistant form with a step of one day, which we can do using an interpolation with smoothing, for example,
Gaussian filter or spline smoothing. We used the first method applying formula (2). Here, we must
choose the optimum smoothing parameter~$a$.

The difficulty is that the ZM2 and NEOS models
use different, but quite similar, initial data and different
 smoothing parameters. Therefore, by varying
the smoothing parameter for the IVS series when
constructing the reference series for comparison, we
can artificially bring this reference series closer to the
series being compared, which will distort our estimates
 of the accuracy of the CPO predictions.

To make our comparison as objective as possible, we calculated the prediction accuracy for six different
smoothing parameters: $a$ = 1, 2, 4, 8, 16, and 32. These calculations give similar results. Figure~3
presents the calculations for two smoothing parameters $a$=2, which is close to the NEOS series, and
$a$=16, which is close to the ZM2 model. The original and smoothed IVS series are shown to the left, and
the rms errors of the predictions to the right. These errors are called external, since they are calculated
relative to the ``external'' reference CPO series. The different smoothing parameters applied to the IVS
series before comparison with the predictions result only in different smoothness of the curves, with no
significant changes in the accuracy of the predictions obtained with the different methods. For completeness
of the comparison, Fig.~4 shows the maximum prediction errors for $a$=16.

\begin{figure*}
\centering
\includegraphics[width=\textwidth,clip]{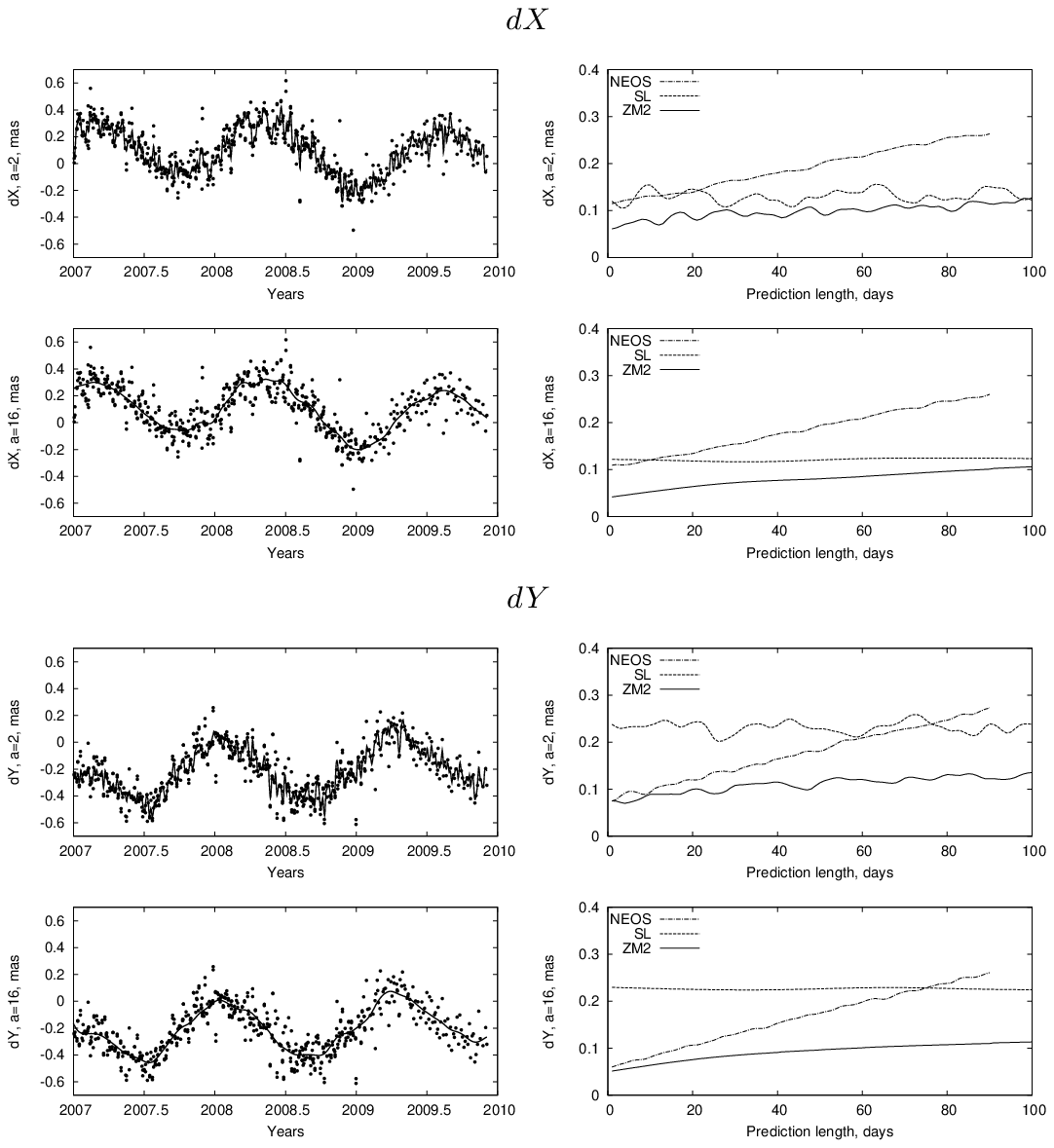}
\caption{External errors of the predictions obtained for various CPO models and two smoothing parameters for the IVS data.
The original and smoothed IVS series are shown to the left, and the rms prediction errors to the right.}
\label{fig:pred_ivs}
\end{figure*}

\begin{figure}[ht]
\centering
\includegraphics[width=0.6\columnwidth,clip]{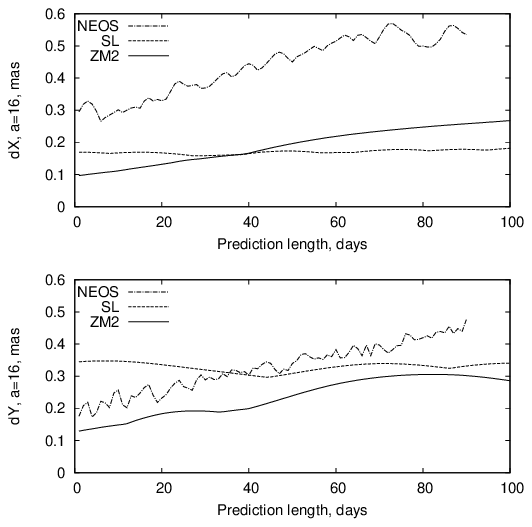}
\caption{Maximum errors of the predictions obtained for various CPO models.}
\label{fig:preda_ivs_max}
\end{figure}

We arrive at the following conclusions about the accuracy of the CPO predictions:
\begin{itemize}
\item the ZM2 model displays the highest prediction accuracy;
\item the NEOS model demonstrates a high accuracy for short-term forecasting, but the accuracy reduces with the prediction length;
\item the error of the SL CPO prediction is mainly determined by the systematic difference between the SL model and the IVS data.
\end{itemize}

Note that, in contrast to the other two models, the Lambert model is an FCN model and not a CPO
model, although it is recommended by the IERS Conventions as a CPO model (which is why we have
included it in our comparison). For this reason, the SL model does not perform well compared to the
NEOS and ZM2 models. Therefore, it is also of interest to study the accuracy of the predictions obtained
using various methods for a given model. For this purpose, we repeated our analysis, but comparing the
predictions with the final series obtained from the same model (instead of comparing with the final IVS
series). The rms prediction errors for this case are shown in Fig. 5. These errors can be considered to be internal.

\begin{figure}[ht]
\centering
\includegraphics[width=0.6\columnwidth,clip]{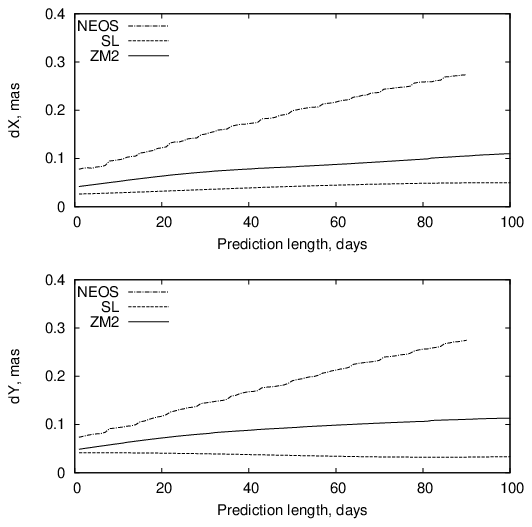}
\caption{Internal errors of the predictions obtained for various CPO models.}
\label{fig:pred_model}
\end{figure}

To estimate the efficiency of the prediction methods, it was proposed in [1] to use the index $P$ (predictability), defined as
\begin{equation}
P(\tau) = 1 - \frac{\sigma_p(\tau)}{\sigma_s} \,,
\label{eq:effectiveness}
\end{equation}
where $\tau$ is the prediction length, $\sigma_p(\tau)^2$ the variance of the prediction errors for the given length,
and $\sigma_s^2$ the variance of the signal predicted. $P=1$ indicates that the prediction coincides with the final results;
this is typical for deterministic processes that are described completely by the model used. $P<0$ indicates
that the prediction is completely inappropriate. In our case, the predicted signal is the CPO series. In general,
the variance of the CPO series depends slightly on the time interval considered. For the total IVS series obtained
from 1984 to 2009, $\sigma_s$=0.18 milliarcsecond (mas) for $dX$ and $\sigma_s$=0.23~mas for $dY$.
For the three last years, $\sigma_s$=0.20~mas for $dX$ $\sigma_s$=0.27~mas for $dY$. An increase in $\sigma_s$
calculated over the last years is not surprising, since this variance is calculated for the original series including
the trend that increases with time. For $\sigma_p(\tau)$, we use the rms prediction errors (see Fig.~3). Figure~6
presents the efficiency of the predictions.

\begin{figure}[ht]
\centering
\includegraphics[width=0.6\columnwidth,clip]{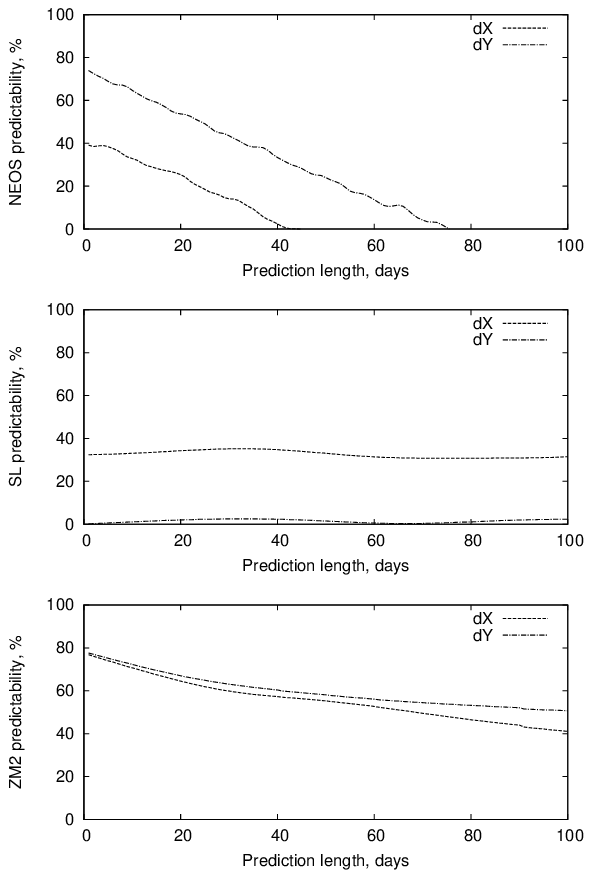}
\caption{Prediction efficiency for the three CPO models.}
\label{fig:effectiveness}
\end{figure}

\section{Summary}

We have analyzed the accuracy of CPO predictions  calculated using three models that are currently
available for practical use: the NEOS model calculated  at the US Naval Observatory, which is part of an
online summary IERS EOP series; the SL model calculated  at the Paris Observatory and recommended
by the IERS Conventions (2003) for high-accuracy modeling of the CP motion; and the ZM2 model
calculated at the Pulkovo Observatory. The analysis used CPO predictions calculated in 2007-2009 and
an a posteriori comparison of these predictions with the final IVS CPO series. Our statistical analysis
of the differences obtained determined the following four characteristics of the accuracy of the prediction
calculated using each method:
\begin{itemize}
\item the rms error of the CPO prediction,
\item the maximum error of the CPO prediction,
\item the efficiency of the CPO prediction,
\item the rms error of the prediction compared with the same model.
\end{itemize}

The first three errors describe the accuracy of the CPO prediction, while the last indicates the accuracy
 of the model self-reproduction. The ZM2 model demonstrates the highest CPO-prediction accuracy.
In addition, only this model is efficient in forecasting both components of the CP motion, $dX$ and $dY$,
over a total prediction length reaching 100 days. The rms error of the one-month ZM2 prediction is about
75-100~$\mu$as, depending on the smoothing parameter used. Since the celestial pole offset, which is the unpredicted
part of the CP motion, reaches 400~$\mu$as and increases with time, this method improves the accuracy
of the CPO prediction in ephemeris calculations by at least a factor of five.

Only two of the models compared---the NEOS and ZM2 models---describe the CP motion completely,
i.e. including the celestial pole offset. The model of Lambert is an FCN model; i.e., it includes
only one CPO component [see (1)], and cannot compete  with the two other models in modeling the total
CP motion, although it can be quite efficient in FCN predictions.

Our main conclusions on the comparative accuracy  of various predictions of the CP motion via a
comparison of the rms prediction errors are supported by our analysis of the prediction efficiency (see Fig.~6).

It is important that we have analyzed here predictions  that are regularly calculated and freely accessible.
Therefore, the results obtained can be helpful for both comparative studies of various predictions of the
CP motion and estimates on the accuracy of the data obtained, as well as for choosing the most appropriate
models for particular applications.

\end{document}